\documentclass[cpl,aps,twocolumn,superscriptaddress]{revtex4-1}

\usepackage{epsfig}
\usepackage{natbib}
\usepackage{mathrsfs}
\usepackage{amsmath}
\usepackage{array}

\begin{document}

\title{Ionic-liquid-gating induced protonation and superconductivity in FeSe, FeSe$_{0.93}$S$_{0.07}$, ZrNCl, 1$T$-TaS$_2$, and Bi$_2$Se$_3$}

\author{Yi Cui}
\thanks{These authors contributed equally to this study.}
\affiliation{School of Mathematics and Physics, North China Electric Power University, Beijing, 102206, China}
\affiliation{Department of Physics, and Beijing Key Laboratory of Opto-electronic Functional Materials $\&$ Micro-nano Devices, Renmin University, Beijing 100872, China}

\author{Ze Hu}
\thanks{These authors contributed equally to this study.}
\affiliation{Department of Physics, and Beijing Key Laboratory of Opto-electronic Functional Materials $\&$ Micro-nano Devices, Renmin University, Beijing 100872, China}

\author{Jin-Shan Zhang}
\email{zhangjs@ncepu.edu.cn}
\affiliation{School of Mathematics and Physics, North China Electric Power University, Beijing, 102206, China}

\author{Wen-Long Ma}
\affiliation{International Center for Quantum Materials, School of Physics, Peking University, Beijing 100871, China}

\author{Ming-Wei Ma}
\affiliation{International Center for Quantum Materials, School of Physics, Peking University, Beijing 100871, China}

\author{Zhen Ma}
\affiliation{National Laboratory of Solid State Microstructures and Department of Physics, Nanjing University, Nanjing 210093, China}

\author{Cong Wang}
\affiliation{Department of Physics, and Beijing Key Laboratory of Opto-electronic Functional Materials $\&$ Micro-nano Devices, Renmin University, Beijing 100872, China}

\author{Jia-Qiang Yan}
\affiliation{Materials Science and Technology Division, Oak Ridge National Laboratory, Oak Ridge, Tennessee 37831, USA}

\author{Jian-Ping Sun}
\affiliation{Beijing National Laboratory for Condensed Matter Physics and Institute of Physics, Chinese Academy of Sciences, Beijing 100190, China}

\author{Jin-Guang Cheng}
\affiliation{Beijing National Laboratory for Condensed Matter Physics and Institute of Physics, Chinese Academy of Sciences, Beijing 100190, China}

\author{Shuang Jia}
\email{gwljiashuang@pku.edu.cn}
\affiliation{International Center for Quantum Materials, School of Physics, Peking University, Beijing 100871, China}
\affiliation{Collaborative Innovation Center of Quantum Matter, Beijing 100871, China}

\author{Yuan Li}
\affiliation{International Center for Quantum Materials, School of Physics, Peking University, Beijing 100871, China}
\affiliation{Collaborative Innovation Center of Quantum Matter, Beijing 100871, China}

\author{Jin-Sheng Wen}
\affiliation{National Laboratory of Solid State Microstructures and Department of Physics, Nanjing University, Nanjing 210093, China}
\affiliation{Innovative Center for Advanced Microstructures, Nanjing University, Nanjing 210093, China}

\author{He-Chang Lei}
\affiliation{Department of Physics, and Beijing Key Laboratory of Opto-electronic Functional Materials $\&$ Micro-nano Devices, Renmin University, Beijing 100872, China}

\author{Pu Yu}
\affiliation{State Key Laboratory of Low Dimensional Quantum Physics and Department of Physics, Tsinghua University, Beijing 100084, China}
\affiliation{Collaborative Innovation Center of Quantum Matter, Beijing 100871, China}

\author{Wei Ji}
\affiliation{Department of Physics, and Beijing Key Laboratory of Opto-electronic Functional Materials $\&$ Micro-nano Devices, Renmin University, Beijing 100872, China}

\author{Wei-Qiang Yu}
\email{wqyu\_phy@ruc.edu.cn}
%\email{wqyu\_phy@ruc.edu.cn;Wqyu.phy<wqyu.phy@gmail.com}
\affiliation{Department of Physics, and Beijing Key Laboratory of Opto-electronic Functional Materials $\&$ Micro-nano Devices, Renmin University, Beijing 100872, China}

\begin{abstract}

We report protonation in several compounds by an ionic-liquid-gating method, under optimized gating conditions. This leads to single superconducting phases for several compounds. Non-volatility of protons allows post-gating magnetization and transport measurements. The superconducting transition temperature $T_{\rm c}$ is enhanced to 43.5\,K for FeSe$_{0.93}$S$_{0.07}$, and 41\,K for FeSe after protonation. Superconducting transitions with $T_{\rm c} \sim 15$\,K for ZrNCl, $\sim$7.2\,K for 1$T$-TaS$_2$, and $\sim$3.8\,K for Bi$_2$Se$_3$ are induced after protonation. Electric transport in protonated FeSe$_{0.93}$S$_{0.07}$ confirms high-temperature superconductivity. Our $^{1}$H nuclear magnetic resonance (NMR) measurements on protonated FeSe$_{1-x}$S$_{x}$ reveal enhanced spin-lattice relaxation rate $1/^{1}T_1$ with increasing $x$, which is consistent with the LDA calculations that H$^{+}$ is located in the interstitial sites close to the anions.

\end{abstract}

\maketitle
Carrier doping is an effective method for tuning metal-insulator transitions and superconductivity. In addition to chemical substitution, electric gating also emerged as an efficient method for tuning carrier density in thin films.~\cite{Ahn_RMP_2006,Ueno_NM_2008,Saito_Science_2015} With the development of various room-temperature ionic liquids, the transistor-like gating method~\cite{Ye_NM_2010,Bollinger_Nature_2011,Ye_Science_2012,Li_Nature_2015,Lu_Science_2015,Miyakawa_PRM_2018} was found to induce a large carrier density for thin films or crystal flakes, through charge polarization. Lithium doping by ionic solid gating was also found to enhance the superconducting transition temperature of thin flakes of FeSe~\cite{ChenXH_PRB_2017}. Recently, tuning of proton or oxygen concentration, using ionic-liquid-gating as a medium, was introduced to modify the lattice structure and magnetism of SrCoO$_{2.5}$~\cite{Yu_nature_2017}. This H$^+$ implantation method was later applied in iron-based superconductors, to induce superconductivity or enhance the superconducting transition temperature in bulk crystals due to an electron doping effect~\cite{Cui_SciBull_2018}.

It is important to note that H$^+$ originates from water contamination in the ionic liquid by this method~\cite{Yu_nature_2017}. The advantages of this technique are that H$^+$ is nonvolatile and the gating is performed near the ambient conditions, which allow various post-gating measurements. However, multiple superconducting phases in protonated FeSe$_{1-x}$S$_x$ emerge, indicating that proton concentration is inhomogeneous across the bulk crystals. The magnetization data in the protonated FeSe$_{0.93}$S$_{0.07}$ show that the volume fraction of the superconducting phase is very low under the reported gating conditions~\cite{Cui_SciBull_2018}.

In this Letter, we report our optimized protonation conditions with this ionic-liquid-gating method, to improve the superconducting volume ratio and the doping homogeneity. The best protonation temperature is found to be 350\,K (higher than the room temperature), with a gating period of 12 days. For FeSe$_{0.93}$S$_{0.07}$, the superconducting volume ratio is largely enhanced compared to the room temperature gating, as determined by the magnetization measurement. Transport measurement is also succeeded to confirm superconductivity. We also apply the optimized protonation on various layered compounds, including FeSe, insulating ZrNCl, 1$T$-TaS$_2$, and Bi$_2$Se$_3$, where protonation either induces superconductivity or enhances the $T_{\rm c}$ largely. In particular for 1$T$-TaS$_2$, we achieve a $T_{\rm c}$ higher than the regular gating method.

\begin{figure}
\includegraphics[width=7cm, height=4cm]{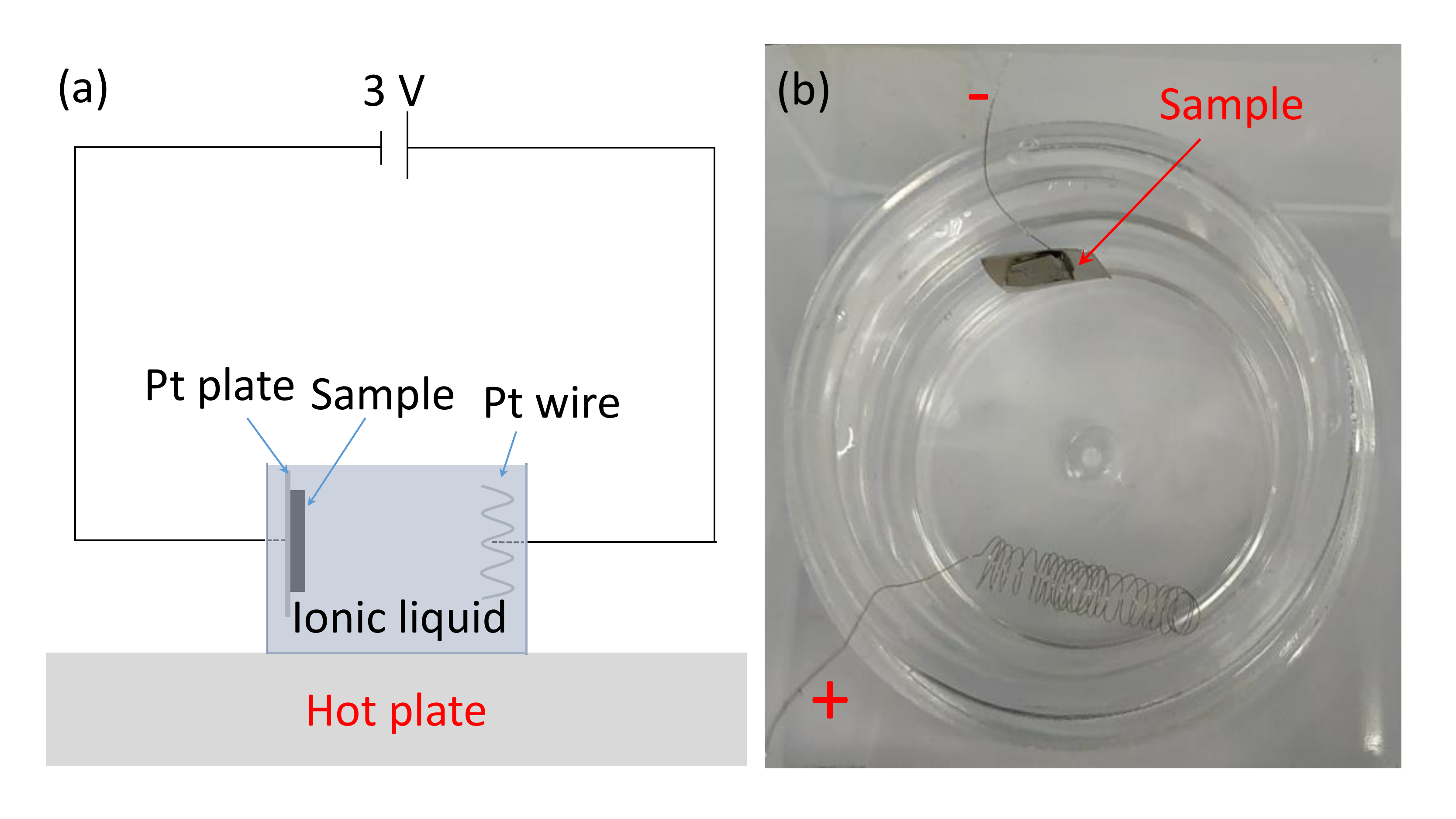}
\caption{\label{fig1} (a) An illustration of the protonation setup.
Platinum electrodes are placed in a container filled with the ionic liquid.
The gating voltage is set to be about 3 V.
The ionic liquid are heated up to 350 K by a hot plate.
(b) A picture of the positive and negative of platinum electrodes, with the
sample attached on the negative electrode.}
\end{figure}

In our experiment, pristine FeSe and FeSe$_{0.93}$S$_{0.07}$ single crystals were made by the vapor transport method~\cite{Bohmer_PRB_2013,Shibauchi_PNAS_2016}. ZrNCl powders were made by the high-pressure synthesis~\cite{ChenX_JPCM_2002}. The 1$T$-TaS$_2$ single crystal was grown by the chemical vapor transport method~\cite{Kuwabara_1986}. Bi$_2$Se$_3$ was grown by the flux method~\cite{Sultana_JSNM_2017}. FeS single crystal was made by the hydrothermal method~\cite{Borg_PRB_2016}. We employ the protonation technique as illustrated in Fig.~\ref{fig1}. As shown in Fig.~\ref{fig1}, samples are attached to the negative electrodes, and a voltage of 3.0\,V is applied as the gating voltage. The ionic liquid EMIM-BF4 is used. The gating temperature is optimized to be 350\,K, which improves proton diffusion efficiency in the crystal. Typical gating period is 12 days when water is nearly fully electrolyzed. The dc magnetization is measured with a magnetic property measurement system (MPMS), and the transport is measured with a physical property measurement system (PPMS). These measurements were successfully performed after gating was removed at the room temperature, which indicates nonvolatile protons are inserted, in contrast to conventional ionic-liquid gating where gating cannot be removed during measurements. The proton NMR is performed by the spin-echo method, and the spin-lattice relaxation rate $1/^1T_1$ is measured by the inversion recovering method.

In the following, we present protonation measurements on these compounds.

\begin{figure}
\includegraphics[width=7.5cm, height=6.5cm]{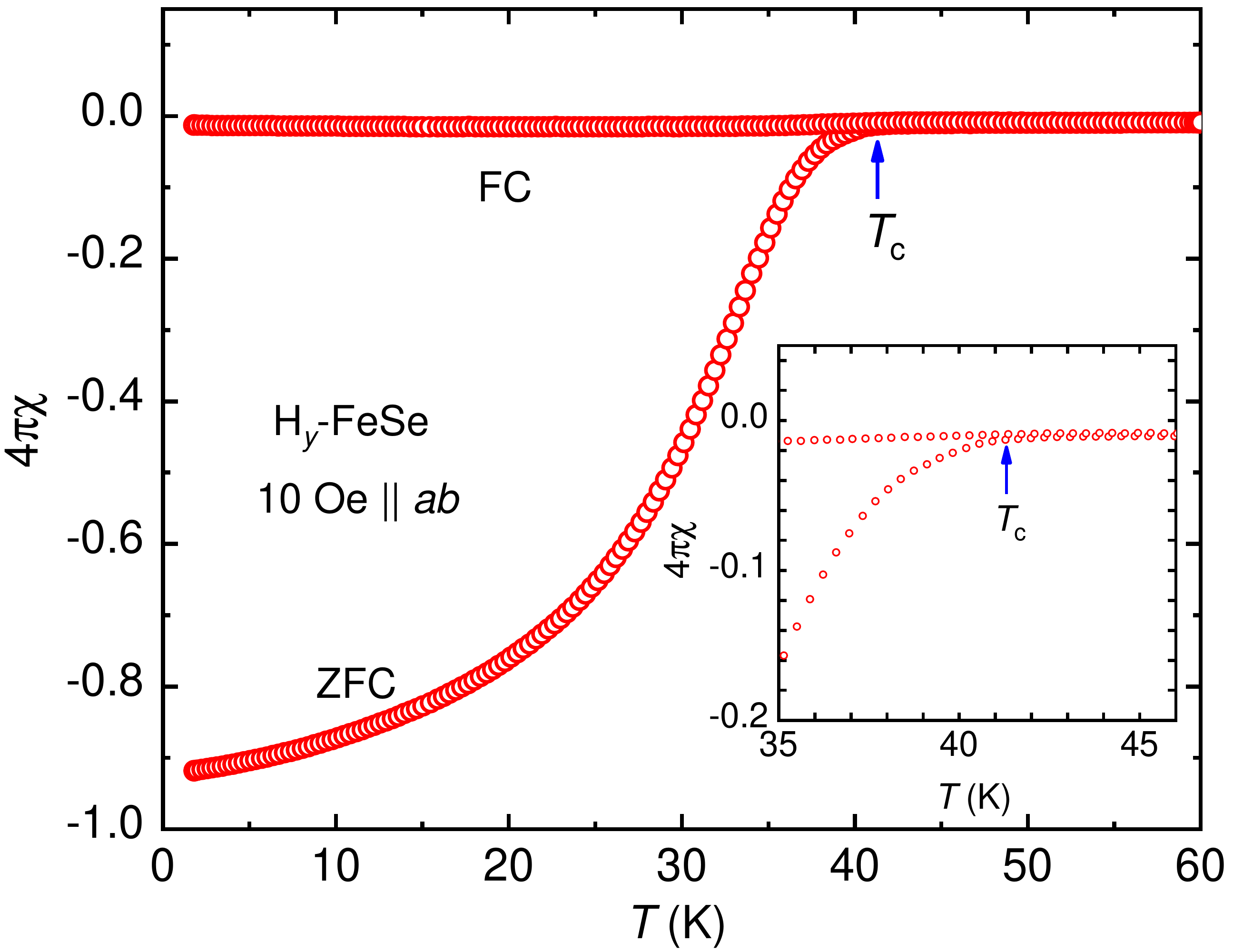}
\caption{\label{fig2} The dc susceptibility of a protonated FeSe
single crystal (size 5mm*5mm*1mm) measured under the field-cooled (FC)
and zero-field-cooled (ZFC) conditions with a magnetic field of 10 Oe.
The arrows points at the superconducting
transition. Inset: An enlarged view of the susceptibility data
close to $T_c$.
}
\end{figure}

{\it FeSe.} Recently, FeSe has attracted a great deal of research attention because of its highly
tunable superconductivity. Its $T_c$ is enhanced from 8.5 K to above 40 K under
high pressure~\cite{Felser_NatMater_2009,sun_NC_2016}, by chemical
intercalation~\cite{ChenXL_PRB_2010,Hatakeda_JPSJ_2013,Dong_JACS_2015,ChenXH_NM_2015},
by ionic-liquid/solid
gating~\cite{ChenXH_PRL_2016,ChenXH_PRB_2017}, or by dimensional reduction into
a single-layer phase~\cite{XueQK_CPL_2012}.

Fig.~\ref{fig2} shows the dc susceptibility $\chi$(T) of a protonated FeSe
single crystal. A rapid drop of $\chi$ are clearly at 41 K seen,
indicating the onset of superconductivity. Therefore, the $T_c$ of FeSe is also largely enhanced by the protonation
technique.

\begin{figure}
\includegraphics[width=8.5cm,height=5cm]{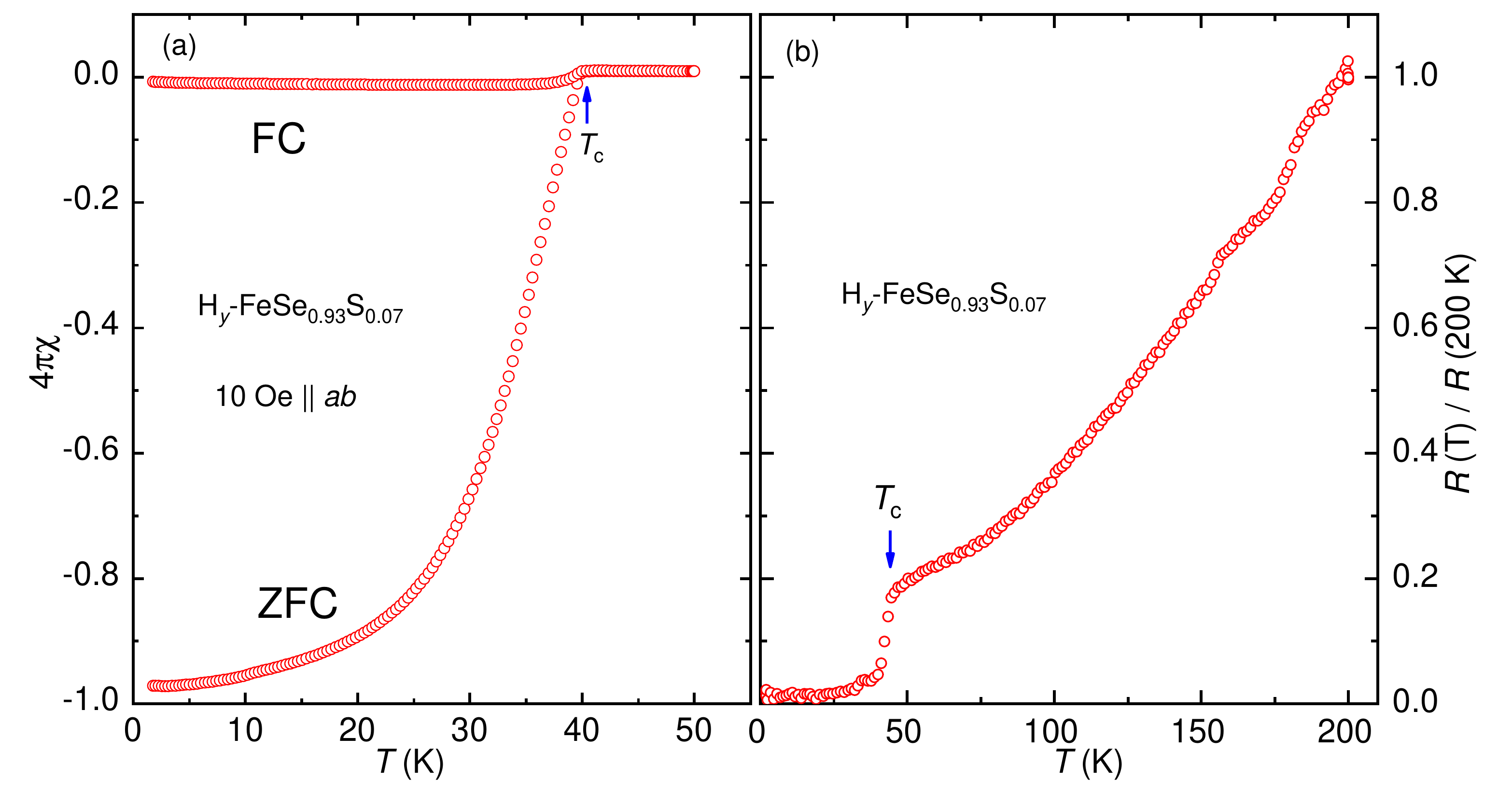}
\caption{\label{fig3} (a) The dc susceptibility of a H$_y$-FeSe$_{0.93}$S$_{0.07}$
single crystal as a function of temperature, measured under ZFC and FC conditions.
(b) The resistance of the crystal as a function of temperature. The arrows mark the onset temperature of superconductivity.}
\end{figure}

{\it FeSe$_{0.93}$S$_{0.07}$.} FeSe$_{1-x}$S$_x$ is a series of compounds, whose $T_c$ ranges between 8 K
 and 13 K for x$<$0.12~\cite{Takano_JPSJ_2009}. Previously, two superconducting transitions, at 25 K
 and 42.5 K were reported in the protonated sample. Here we show that with
 increased protonation temperature at 350 K, a single high-$T_c$ phase is realized.
 Figure~\ref{fig3} shows the dc susceptibility $\chi$(T) and the resistance data R(T)
 of a protonated FeSe$_{0.93}$S$_{0.07}$ single crystal. The susceptibility data shows $T_c$~$\approx$ 41 K,
 seen by the drop of $\chi$ (Fig. 3 (a)). By contrast, the resistance
 data shows a higher onset $T_c$ of 43.5 K as indicted in Fig.~\ref{fig3}, by a sudden drop of resistance upon cooling.

\begin{figure}
\includegraphics[width=8.5cm,height=7cm]{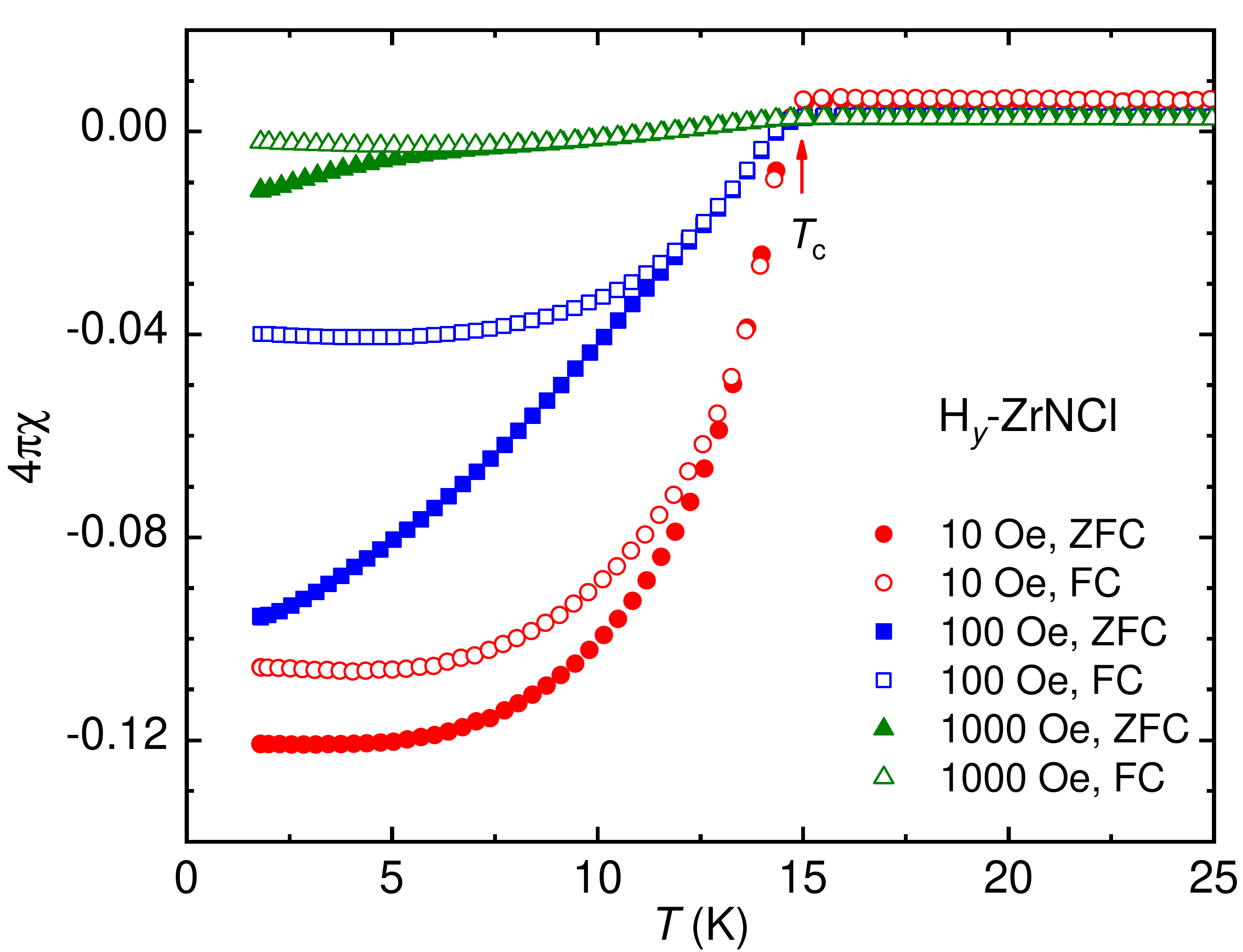}
\caption{\label{fig4} The dc susceptibility $\chi$ of H$_y$-ZrNCl pellets,
measured under FC and ZFC condition with different fields.
The arrow marks the onset temperature of superconductivity.}
\end{figure}

{\it ZrNCl.} ZrNCl is a layered material with electric gating or lithium doping. Superconductivity can be induced in ZrNCl by electric gating or lithium doping~\cite{Ye_NM_2010,Saito_Science_2015,Taguchi_PRL_2006}. We pressed ZrNCl powders into thin pellets and then doped H$^+$ with the current ionic-liquid-gating method. The samples turn from blue into black upon proton doping. Figure 4 shows the dc susceptibility of the proton-doped ZrNCl. The sharp drop of $\chi$ below 15\,K shows the onset of superconductivity, with field up to 1000\,Oe. The volume ratio of the superconducting phase, estimated from the ZFC data at 10\,Oe field, is about 12$\%$. This suggests that proton doping is very efficient. We note that an ionic-liquid gating on ZrNCl at low temperatures is also reported, which proposes that the depletion of Cl$^{-}$ concentration causes superconductivity~\cite{zhangshuai}.

\begin{figure}
\includegraphics[width=8.5cm,height=7cm]{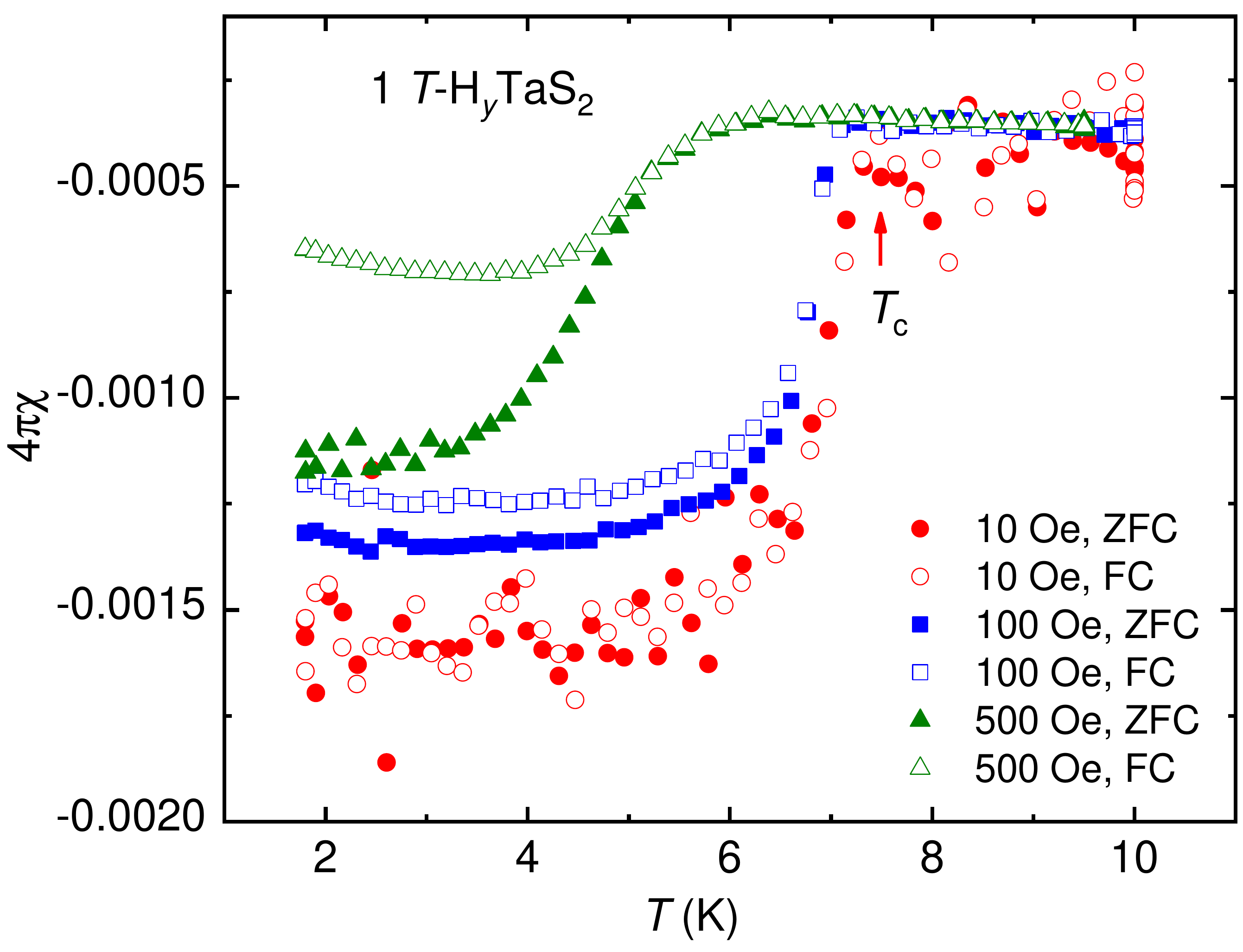}
\caption{\label{fig5} The dc susceptibility $\chi$ of a protonated 1{\it T}-TaS$_2$ single crystal,
measured under FC and ZFC condition, under various fields. The arrow marks the
onset of the superconducting transition.}
\end{figure}

{\it 1T-TaS$_2$.} 1$T$-TaS$_2$ is a layered compound with a triangular lattice. It goes through a series of charge-density-wave (CDW) transitions upon cooling~\cite{Wilson_1975,Thomson_1994}. Superconductivity can be achieved by chemical doping, where the highest $T_{\rm c}$ is achieved at 3.5\,K~\cite{Liu_APL_2013}. Here we performed protonation on 1$T$-TaS$_2$ single crystals. The dc magnetization of a protonated sample is shown in Fig.~\ref{fig5}, measured under FC and ZFC conditions at different fields. The superconducting transition temperature $T_{\rm c}$ is found to be $\sim$7.2\,K under 10\,Oe field. We note that this transition temperature is higher than that achieved by the chemical doping. With an applied field of 500\,Oe, the superconducting transition is still observed.

\begin{figure}
\includegraphics[width=8.5cm,height=7cm]{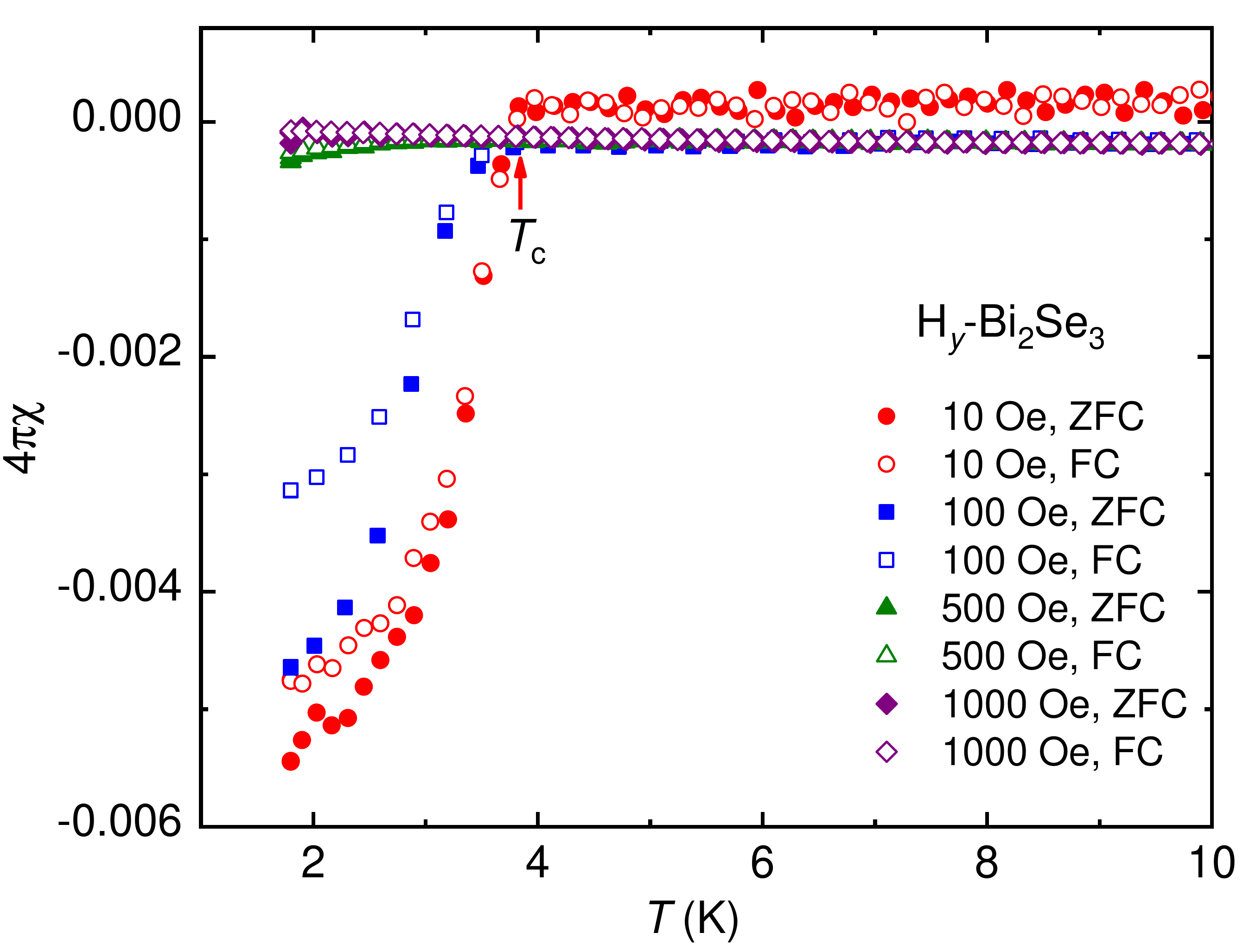}
\caption{\label{fig6} The dc susceptibility $\chi$ of a protonated Bi$_2$Se$_3$
single crystal, measured under FC and ZFC condition at different fields.
The arrow marks the onset of superconducting tansition at $T_C$.}
\end{figure}

{\it Bi$_2$Se$_3$.} As a topological insulator, Bi$_2$Se$_3$ has caused a lot of research interests~\cite{zhang_NP_2009,Xia_NP_2009}. Superconductivity can be achieved upon Cu or Sr doping into this material~\cite{Hor_PRL_2010,Liu_JACS_2015}. Here we find that by protonation, superconductivity can also be achieved. As seen in Fig.~\ref{fig6}, the superconducting transition temperature $T_{\rm c}$ is found to be 3.8\,K, which is close to that reported by the chemical doping. With an applied field of 500\,Oe, superconductivity is highly suppressed. Since protonation does not induce chemical substitution, our study indicates that chemical doping in the interstitial sites is important for the occurrence of superconductivity. Further studies on the protonation of induced superconductivity in this compound, regarding to possible topological superconductivity, are demanded.

\begin{figure}
\includegraphics[width=8.5cm,height=7cm]{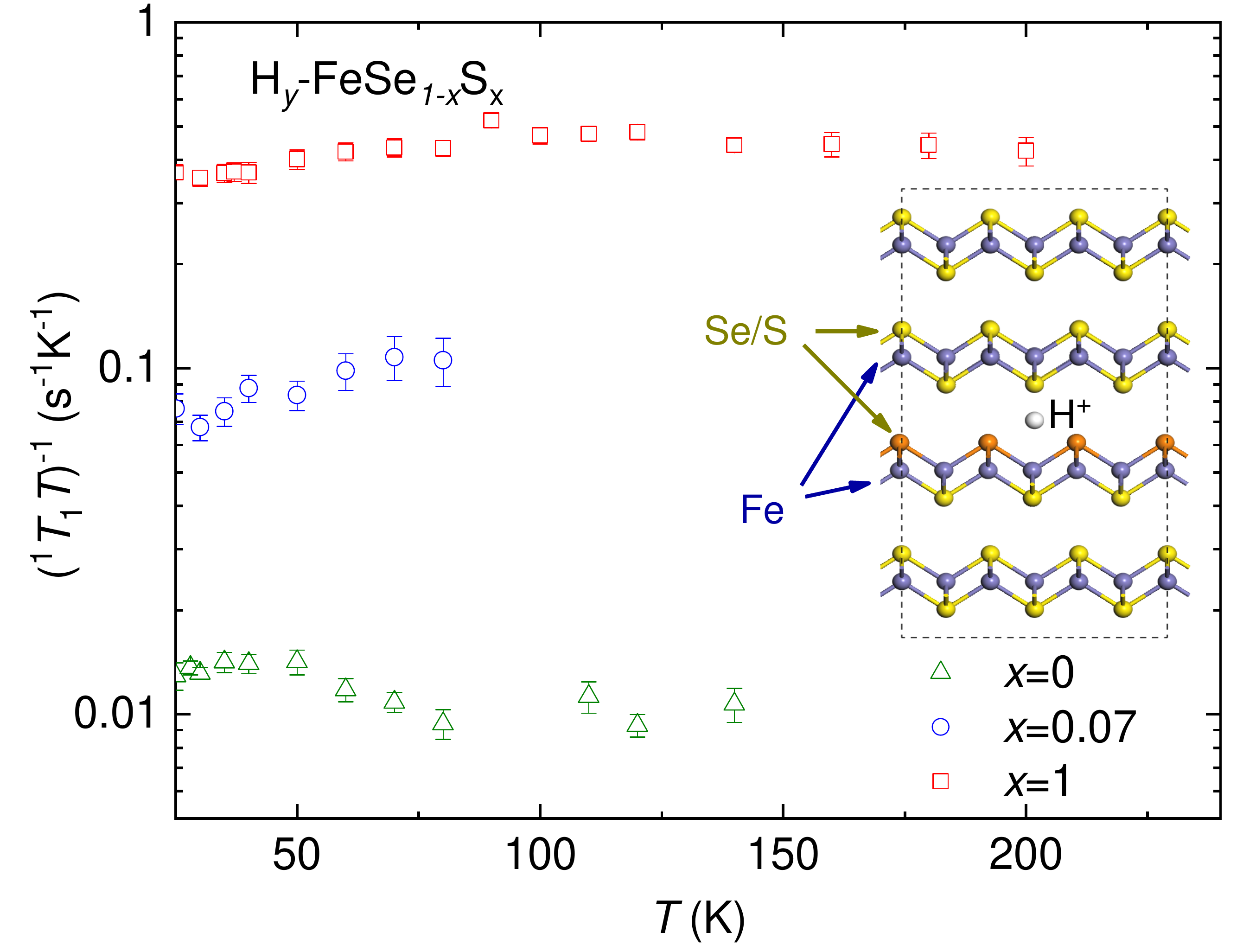}
\caption{\label{fig7} The $^1$H spin-lattice relaxation rate divided
by temperature $1/^1T_1T$ of H$_y$FeSe$_{1-x}$S$_x$ single crystals, measured with a
magnetic field of 5 T along the $c$-axis. Inset: The proton position in the lattice obtained
by the LDA calculations.}
\end{figure}

$^1$H NMR studies on H$_y$FeSe$_{1-x}$S$_{x}$: In H$_y$FeSe$_{1-x}$S$_{x}$ compounds, intrinsic $^1$H NMR spectra was observed. Figure 7 shows the $^1$H spin-lattice relaxation rates divided by temperatures, versus temperature for protonated FeSe$_{1-x}$S$_x$ with $x$=0, 0.07 and 1. Above 50\,K, $1/^1T_1T$ stays nearly constant but varies with $x$, which indicate that doped protons are detected by the current measurements. Indeed, the increase of $1/^{1}T_1$ with increasing $x$ is consistent with the LDA calculations that H$^+$ is inserted in the interstitial sites as discussed below. Since the $c$-axis lattice parameter is reduced with increasing $x$~\cite{Takano_JPSJ_2009}, the hyperfine coupling between $^1$H and the FeSe plane increases with increasing S$^{2-} $ concentration.

Our LDA calculations indicate that H$^+$ is located in the interstitial sites close to the anion Se$^{2-}$/S$^{2-}$, as shown by the schematic drawing in the inset of Fig.\,7. This can be understood as an effect of coulomb attraction between H$^{+}$ and Se$^{2-}$/S$^{2-}$. So far, we have found that this doping method is efficient in layered compounds, which indicates that H$^+$ is most likely doped between the layers as in H$_y$FeSe$_{1-x}$S$_{x}$.

{\it Discussions and summary.} Our XRD measurement did not resolve the change of lattice structure after protonation, which suggests that the chemical pressure effect of proton insertion is possibly very small. As a result, an electron-doping should be primarily responsible for the change of $T_{\rm c}$.

\begin{table}
\label{t1}
\caption{$T_C$ of the materials before and after protonation}
  \centering
\begin{tabular}{m{2cm}<{\centering}|c|c|c|c|c}
\toprule
  Compound & FeSe & FeSe$_{0.93}$S$_{0.07}$  & ZrNCl & 1{\it T}-TaS$_2$ & Bi$_2$Se$_3$    \\
\hline
  $T_c$ before protonation &9 K & 8 K & 0 & 0 & 0   \\
\hline
  $T_c$ after protonation & 41 K & 43.5 K & 15 K & 7.2 K & 3.8 K   \\
\botrule
\end{tabular}
\end{table}

In Table 1, we summarize all the $T_{\rm c}$ of different compounds, before and after protonation under the current optimized conditions. The optimization at 350\,K suggests that the efficiency of proton doping is caused by a balance between proton diffusion into the sample and the evasion out of the sample, both of which increase with temperature. The current method supplies a universal electron doping method, which could be widely used in tuning and searching for superconductivity and metal-insulator transitions in the layered compounds.

Work at RUC was supported by the National Natural Science Foundation of China with Grand Nos.~51872328, 11622437, 11574394, 11774423, and 11822412,
the Strategic Priority Research Program of Chinese Academy of Sciences (Grant No. XDB30000000),
the Ministry of Science and Technology of China with Grand No.~2016YFA0300504,
the Fundamental Research Funds for the Central Universities, and the Research Funds of Renmin University of China (RUC) (15XNLQ07, 18XNLG14, 19XNLG17).
SJ was supported by the National Natural Science Foundation of China with Grand Nos.~11774007 and U1832214.
YC was supported by the Outstanding Innovative Talents Cultivation Funded Programs 2018 of Renmin University of China.
JQY was supported by the U.S. Department of Energy, Office of Science, Basic Energy Sciences, Division of Materials Sciences and Engineering.

\end{document}